\documentclass[10pt,letterpaper]{article}
\usepackage{opex3}

\newcommand{\de}{\partial}

\newcommand{\eq}[2]{\begin{equation} \label{#1} #2 \end{equation}}

\newcommand{\sgn}{\textrm{sgn}}

\begin{document}

\title{Mimicking the nonlinear dynamics of optical fibers with waveguide arrays: towards a spatiotemporal supercontinuum generation}

\author{Truong X. Tran$^{1,2}$ and Fabio Biancalana$^{2,3}$}

\address{$^{1}$Dept. of Physics, Le Quy Don University, 236 Hoang Quoc Viet Str., 10000 Hanoi, Vietnam\\
$^{2}$Max Planck Institute for the Science of Light, G\"{u}nther-Scharowsky-Str. 1/Bau 24, \\ 91058 Erlangen, Germany\\
$^{3}$School of Engineering and Physical Sciences, Heriot-Watt University, \\ EH14 4AS Edinburgh, UK}


\email{truong.tran@mpl.mpg.de} 

\homepage{http://mpl.mpg.de/mpf/php/abteilung3/jrg/index.html} 


\begin{abstract}
We numerically demonstrate the formation of the spatiotemporal version of the so-called diffractive resonant radiation generated in waveguide arrays with Kerr nonlinearity when a long pulse is launched into the system. The phase matching condition for the diffractive resonant radiation that we have found earlier for CW beams also works well in the spatiotemporal case. By introducing a linear potential, one can introduce a continuous shift of the central wavenumber of a linear pulse, whereas in the nonlinear case one can demonstrate that the soliton self-wavenumber shift can be compensated by the emission of diffractive resonant radiation, in a very similar fashion as it is done in optical fibers. This work paves the way for designing unique optical devices that generate spectrally broad supercontinua with a controllable directionality by taking advantage of the combined physics of optical fibers and waveguide arrays. 
\end{abstract}



\section{Introduction}

\label{introduction}

Waveguide arrays (WAs) present a unique discrete periodic photonic platform for exploring many interesting fundamental phenomena such as discrete diffraction \cite{nature,jones}, discrete solitons \cite{nature,christodoulides,kivshar}, and photonic Bloch oscillations \cite{nature,peschel,pertsch2,lenz,lederer}. In applications, 2D networks of nonlinear waveguides with discrete solitons may be useful for designing signal-processing circuits \cite{christodoulides2}. Binary WAs have also been intensively used to mimic relativistic phenomena typical of quantum field theory, such as {\em Zitterbewegung} \cite{zitterbewegung}, Klein paradox \cite{klein}, fermion pair production \cite{fermionpairproduction} and conical diffraction for massless Dirac fields \cite{diracequation}. Quite recently, the optical analogue of relativistic Dirac solitons with exact analytical solutions in binary WAs has been found \cite{truong1} which could potentially pave the way for using binary WAs as a classical simulator of quantum {\em nonlinear} effects arising from the Dirac equation, something that is thought to be impossible to achieve in conventional (i.e. linear) quantum field theory.

The dispersive resonant radiation (DisRR), which emerges due to higher-order dispersion (HOD) terms, has been well explored in the last decade in the temporal case in optical fibers \cite{kuehl,wai,karpman,akhmediev,biancalana}. When an ultrashort pulse is launched into a waveguide, a DisRR due to the phase matching between the fiber and the soliton group velocity dispersion (GVD) generates one or more new frequencies \cite{akhmediev,biancalana}. This DisRR, together with other well-known nonlinear effects such as self- and cross-phase modulation, soliton fission \cite{fission}, and stimulated Raman scattering \cite{srs}, are the main ingredients of the supercontinuum generation (SCG) \cite{dudley,agrawal1}, especially in highly nonlinear photonic crystal fibers \cite{russell,dudley}. SCG has quickly established itself as one of most important phenomena in nonlinear fiber optics which has led to a number of important technological advances in various fields, such as spectroscopy and medical imaging \cite{alfano}, metrology \cite{holzwarth}, and the realization of broadband sources \cite{agrawal2}, just to name a few.

In a recent study the diffractive resonant radiation (DifRR) - an analogue of DisRR - has been found when a spatial soliton in continuous-wave (CW) regime is launched into WAs \cite{truong2}. We have shown in Ref. \cite{truong2} that when the phase matching condition is satisfied, a spatial soliton emits DifRR with a new well-defined direction, i.e. transverse wavenumber. Moreover, due to the periodicity of discrete systems, and thus the existence of a Brillouin zone, the so-called {\em anomalous solitonic recoil} - an unusual effect which cannot exist in continuous media - has been discovered \cite{truong2}.

Inspired by the above recent achievements in the DifRR studies, in this paper we show that DifRRs can occur not only in the CW regime, but also in the spatiotemporal case, i.e. when launching pulses with a finite temporal duration. Thus DifRR is a universal phenomenon occurring in WAs and other nonlinear periodic systems. We also demonstrate that the anomalous compensation of the soliton self-wavenumber shift (SSWS) - an analogue of the Raman soliton self-frequency shift (SSFS) in the single fiber operating in the temporal case - can exist in specially excited WAs.

The structure of the paper is as follows. In section \ref{phasematchingcondition} we briefly review the main results achieved in Ref. \cite{truong2} on the phase matching condition in the CW regime which is fundamental for further parts of this paper. In section \ref{spatiotemporalcase} we numerically show DifRRs generated in the spatiotemporal case. Finally, in section  \ref{SSWS} we give a simple interpretation of the photonic Bloch oscillation by using the concept of wavenumber shift, and then we detail the dynamics of the Raman-like SSWS in WAs in the nonlinear regime. Conclusions are given in section \ref{conclusions}.

\section{Phase matching condition for diffractive resonant radiation in cw regime}
\label{phasematchingcondition}
In this section we briefly review the phase matching condition for DifRRs obtained earlier in Ref. \cite{truong2}. Light propagation in a discrete, periodic array of Kerr nonlinear waveguides can be described, in the CW regime, by the following well-known dimensionless set of ordinary differential equations \cite{kivshar,lederer,agrawal1}:
\eq{CWCM1}{i\frac{da_{n}(z)}{dz} + c[a_{n+1}(z)+ a_{n-1}(z)] + |a_{n}(z)|^{2}a_{n}(z)=0,}
where $a_{n}$ is the electric field amplitude in the $n$-th waveguide, $n=\left\{1,\ldots,N\right\}$, $N$ is the total number of waveguides, $z$ is the longitudinal spatial coordinate, $c$ is the parameter which is related to the coupling coefficient  resulting from the field overlap between neighboring waveguides.

By using the stationary discrete plane wave solution for the $n$-th waveguide $a_{n}(z) = a_{0}\exp[i(nk_{x}d +\kappa_{z}z)]$ one arrives, in the linear case, at the well-known dispersion relation between $\kappa_{z}$ and $k_{x}$ \cite{jones}:
\eq{dispersion}{\kappa_{z}(k_{x})=2c\cos(k_{x}d),} where $d$ is the center-to-center spacing between two adjacent waveguides, and $k_{x}$ is the transverse wavenumber. It is clear from Eq. (\ref{dispersion}) that $\kappa_{z}$ is periodic in $\kappa \equiv k_{x}d$, which represents the phase difference between adjacent waveguides. Thus, within the coupled mode approximation, it suffices to investigate the first Brillouin zone of the folded dispersion, $-\pi \leq \kappa \leq \pi$.

Since a typical input beam has a finite width covering several waveguides, its Fourier spectrum has a certain bandwidth with a central transverse wavenumber $\kappa_{0}$, which is fixed by the input angle of incidence of the exciting beam. We can then use a Taylor expansion of Eq. (\ref{dispersion}) as follows:
\eq{taylor}{\kappa_{z}(\kappa)=\kappa_{z}(\kappa_{0}) + \sum_{m\geq1}\frac{D_{m}}{m!}\Delta\kappa^{m},} where $\Delta\kappa \equiv \kappa - \kappa_{0}$, and $D_{m} \equiv (d^{m}\kappa_{z}/d\kappa^{m})|_{\kappa_{0}}$ is the $m$-th order diffractive Taylor coefficient [thus, $D_{1} = -2c\mathrm{sin}(\kappa_{0})$, $D_{2} = -2c\mathrm{cos}(\kappa_{0})$, etc., all the derivative can obviously be calculated explicitly].  This shape of $D_{2}$ is analogous to the GVD of photonic crystal fibers in the temporal case, which typically exhibits two zero-GVD points \cite{russell}.

By using the approach developed in Refs. \cite{lederer,pertsch}, we approximate the discrete variable $n$ with a continuous one. This is justified {\em a posteriori} since we shall use pulses and solitons that extend for several waveguides. Defining now $n$ as a continuous variable of the distributed amplitude function $\Psi(n,z) = a_{n,z}\exp(-i\kappa_{0}n)$, after dropping the two low-order terms [$\kappa_{z}(\kappa_{0})$ and $-iD_{1}\de_{n}$] by using appropriate transformations one arrives at the following equation, (see Ref. \cite{truong2} for more details):
\eq{mainsimplified}{\left[ i\de_{z} - \frac{D_{2}}{2}\de_{n}^{2} + \sum_{m\geq3}\frac{D_{m}}{m!}(-i\de_{n})^{m} +  |\Psi(n,z)|^{2}\right]\Psi(n,z) =0.}

Equation (\ref{mainsimplified}) is {\em formally identical} to the well-known generalized nonlinear Schr\"odinger equation (GNLSE), which describes the evolution of pulses in a single optical fiber, plus HOD terms \cite{agrawal2}. We should emphasize that in Eq. (\ref{mainsimplified}) we have the transverse spatial variable $n$ instead of the temporal variable $t$ of the conventional GNLSE.

In the temporal version of the GNLSE, it is well-known that a temporal soliton propagating in a fiber emits small-amplitude, dispersive and quasi-monochromatic waves at well-defined frequencies (the DisRR) when the linear fiber dispersion and the nonlinear soliton dispersion are matched \cite{akhmediev,biancalana}. Analogously, in a WA, a {\em spatial} soliton, which in the continuous variables approximation extends over several waveguides, emits during the propagation a similar kind of small-amplitude diffractive radiation, within a narrow wavenumber range, due to the phase-matching between the spatial soliton nonlinear dispersion and the linear array dispersion given by Eq. (\ref{dispersion}). By using the perturbation approach which was developed for DisRRs in Ref. \cite{akhmediev}, we have obtained the phase-matching condition for the DifRR in a similar way \cite{truong2}. We first found the unperturbed soliton solution of Eq. (\ref{mainsimplified}) where all diffractive terms $D_{m\geq3}$ are dropped. Under these conditions, the soliton solution is given by:
\eq{soliton}{a_{\rm sol}(z,n)= A_{0}\mathrm{sech}\left(\frac{nA_{0}}{\sqrt{2c\mathrm{cos}(\kappa_{0})}}  \right)\mathrm{exp}(ik_{\rm sol}z),} where $k_{\rm sol} = A_{0}^{2}/2$. The bright soliton solution (\ref{soliton}) only exists when $2c\mathrm{cos}(\kappa_{0}) > 0$, i.e. only in half of the Brillouin zone, where $-\pi/2 < \kappa_{0}<\pi/2$. Then we looked for the linearized dispersion relation of plane wave solutions of Eq. (\ref{mainsimplified}), by substituting $\mathrm{exp}[i(k_{\rm lin}z + \Delta \kappa n)]$ into Eq. (\ref{mainsimplified}) and using Eq. (\ref{taylor}). We obtained:
\eq{klin}{k_{\rm lin}(\Delta \kappa)\equiv\sum_{m\geq2}\frac{D_{m}}{m!}\Delta\kappa^{m} =2c[\mathrm{cos}(\kappa) - \mathrm{cos}(\kappa_{0}) + \mathrm{sin}(\kappa_{0})\Delta\kappa].} In Eq. (\ref{klin}), $\kappa_{0}$ is the central wavenumber (which is related to the incident angle) of the incident beam, while $\Delta\kappa$ is the detuning from $\kappa_{0}$, and $\kappa = \kappa_{0}+\Delta\kappa$. Energy exchange between radiation and soliton is possible for those values of $\Delta\kappa$ that satisfy
\eq{phasematching}{k_{\rm lin}(\Delta\kappa)=k_{\rm sol},} where $k_{\rm sol}$ is constant and has been defined above. This phase matching condition, an implicit equation for the radiation wavenumber detuning $\Delta\kappa$, is the central result in Ref. \cite{truong2}. It is important to note that although the phase matching condition expressed in Eq. (\ref{phasematching}) has been derived from the continuous model of Eq. (\ref{mainsimplified}), such formula very accurately predicts the DifRR wavenumber in the full original discrete model of Eq. (\ref{CWCM1}), as shown in Ref. \cite{truong2}.

\section{Diffractive resonant radiation in the spatiotemporal case}
\label{spatiotemporalcase}
\begin{figure}[htb]
  \centering \includegraphics[width=0.9\textwidth]{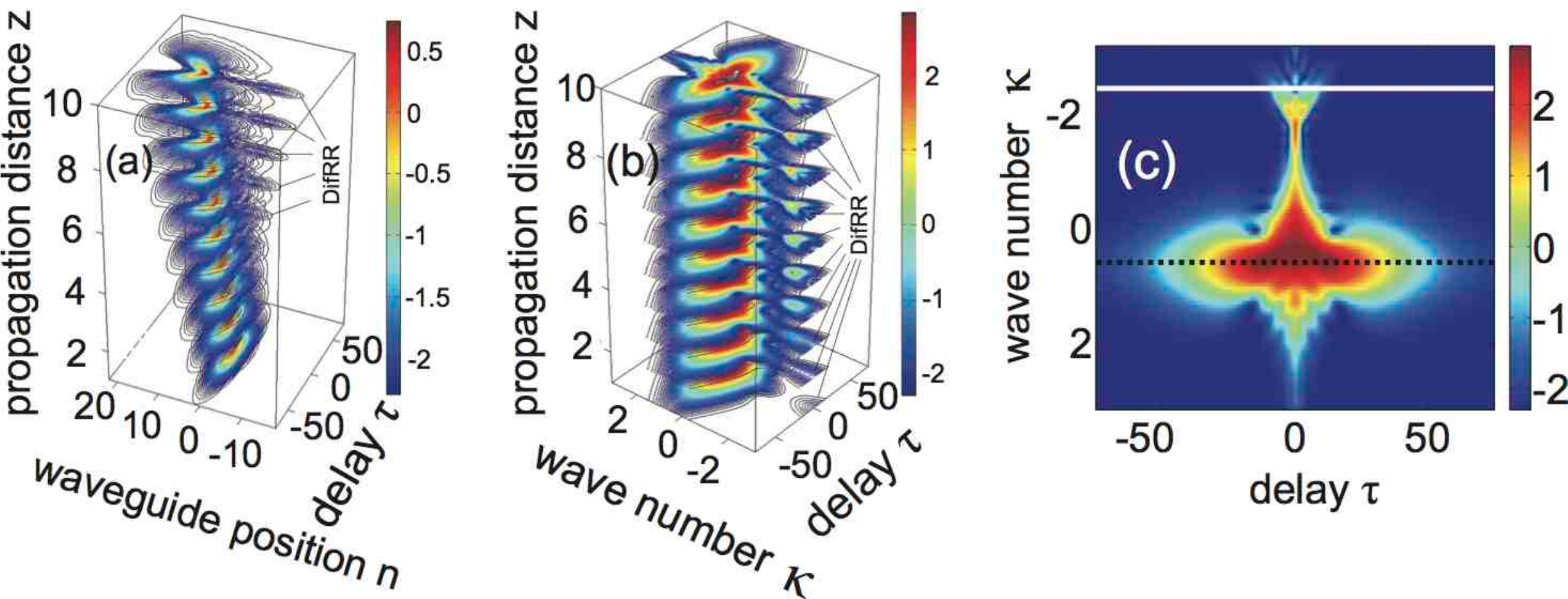}
\caption{\small{(Color online) (a,b) Diffractive resonant radiation in the spatiotemporal case in the ($n,z,\tau$)-space (a) and ($\kappa,z,\tau$)-space (b). (c) The close view of the uppermost slide in Fig. \ref{fig1}(b) is depicted separately with the black dashed line representing the input wavenumber of the soliton ($\kappa_{0}=0.6$) and the solid white line showing the predicted DifRR wavenumber calculated by using the phase matching condition with this input soliton wavenumber value.   Parameters: $c=1$, $\kappa_{0}=0.6$, $A_{0}=1$.}}
  \label{fig1}
\end{figure}
In Ref. \cite{truong2} we have just analyzed the DifRR in the CW regime. In this section we will investigate the DifRR for the spatiotemporal case when a laser pulse is launched into the WA. In this case Eq.(\ref{CWCM1}) should be modified as follows
\eq{CWCMF}{i\de_{z}a_{n} + c [a_{n+1} + a_{n-1}] + \frac{1}{2}\de_{\tau}^{2}a_{n} + |a_{n}|^{2}a_{n}=0,} where $\tau$ is the dimensionless delay variable. In Eq.(\ref{CWCMF}), as proof of principle, we ignore the 3rd and higher-order time derivatives, but these HOD terms and other nonlinear high-order terms such as self-steepening and Raman nonlinearity \cite{agrawal2} can be easily implemented in Eq.(\ref{CWCMF}). The evolution of the pulse is illustrated in Fig.\ref{fig1}(a) in the (n,z,$\tau$)-space. The evolution of the Fourier transform of the field $a(n)$ with respect to the spatial coordinate $n$ of Fig. \ref{fig1}(a) along $z$ is shown in Fig. \ref{fig1}(b) in the ($\kappa$,z,$\tau$)-space. Parameters used for Fig.\ref{fig1} are as follows: $c=1$, $\kappa_{0}=0.6$, $A_{0}=1$, and the input condition $A(n,\tau) = a_{sol}(0,n)\mathrm{exp}(i\kappa_{0}n)\mathrm{sech}(\tau/20)$ where $a_{sol}(0,n)$ has the form of Eq.(\ref{soliton}) at the input ($z$ = 0). As shown in Fig.\ref{fig1}(a) at the input the pulse projection in the ($n,\tau$)-plane has an oval shape which is stretched along the axis $\tau$ (long-pulsed regime). At this stage the pulse projection in the ($\kappa$,$\tau$)-plane also has an oval shape [see Fig.\ref{fig1}(b)]. After some propagation the central part of the pulse with high intensity (around $\tau$ = 0) emits the DifRR, then the projections of the pulse in both (n,$\tau$)- and ($\kappa$,$\tau$)-plane are distorted. In the ($n,\tau$)-plane the oval shape is deformed into a V-shape where two tails of the pulse is temporally split into two sub-pulses together with the DifRR, whereas in the ($\kappa$,$\tau$)-plane the oval shape is deformed into a spade-like shape together with the DifRR generation.

The uppermost slide of the Fig. \ref{fig1}(b) is shown separately in Fig. \ref{fig1}(c) where dashed black horizontal line represents the input wavenumber ($\kappa_{0}=0.6$), while the solid white line is obtained by solving Eq. (\ref{phasematching}) numerically, showing good agreement with the pulse propagation result. Note that there is a small deviation between the final central DifRR wavenumber and the predicted white line. This is due to the fact that the wavenumber of the DifRR represented by the white line in Fig. \ref{fig1}(c) is calculated by using the phase matching condition with the input soliton wavenumber ($\kappa_{0}=0.6$), but during the propagation, the central wavenumber of the soliton slightly decreases towards the value $\kappa=0$, which results in the slight shift of the eventual DifRR wavenumber towards this value due to the phase matching condition [see also Fig. 1(b) in Ref. \cite{truong2} for more details]. Thus, together with the results obtained in Ref. \cite{truong2}, we have shown that the DifRR generation is a quite universal phenomenon in periodic WAs and probably also in other periodic photonic systems under the condition that an input CW beam or long pulse with right profile (soliton) is launched into the WA under any angle large enough compared to the normal incidence. The generation of DifRRs in WAs is as universal as the one of DisRRs in a single optical fiber.

\section{Compensation of the soliton self-wavenumber shift}
\label{SSWS}

\begin{figure}[htb]
\centering\includegraphics[width=10cm]{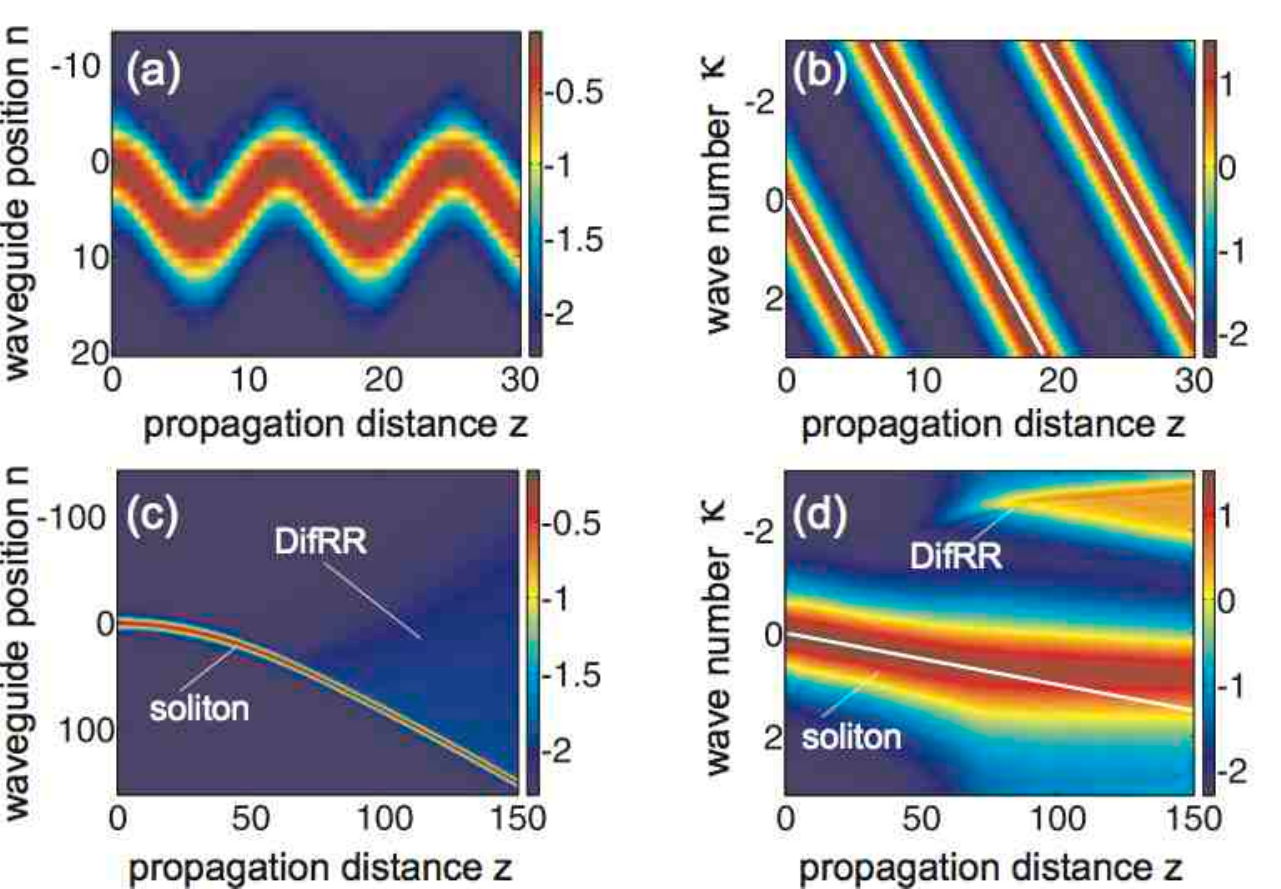}
\caption{\small{(Color online) (a,b) PBOs in the linear regime depicted in the ($n,z$)-plane (a) and ($\kappa,z$)-plane (b). Parameters for (a,b): $c=1$, $\kappa_{0}=0$, $A_{0}=0.8$, $\alpha=0.5$. (c,d) Anomalous compensation of SSWS in the nonlinear regime depicted in the ($n,z$)-plane (c) and ($\kappa,z$)-plane (d). Parameters for (c,d): $c=1$, $\kappa_{0}=0$, $A_{0}=0.6$, $\alpha=0.01$. White straight lines in (b,d) represent the wavenumber calculated by using Eq.(\ref{fSR}).}}
  \label{fig2}
\end{figure}

The Raman soliton self-frequency shift (SSFS) \cite{mitschke,gordon} is a fundamental effect acting on temporally short pulses in silica fibers, and plays a crucial role in SCG \cite{dudley}. In fibers, SSFS leads to a continuous shift of the central frequency $\omega_{0}$ of the soliton towards the red part of the spectrum \cite{mitschke}. When $\omega_{0}$ is sufficiently close to one of the zero-GVD points of the fiber, DisRR is emitted, and the SSFS can be compensated \cite{biancalana,skryabinscience}.

Here we show that an analogue of the Raman SSFS can be obtained in WAs by applying an external linear potential across the transverse coordinate $n$, which changes the propagation constant along the array in a linear fashion, by using for instance the electro-optic \cite{peschel} or thermal-optic effect \cite{pertsch2}. In the continuous limit, this results in adding a term $\alpha n a_{n}(z)$ to the left-hand side of Eq. (\ref{CWCM1})
\eq{CWCM2}{\left(i\frac{d}{dz} + \alpha n\right)a_{n} + c [a_{n+1}+ a_{n-1}] + |a_{n}|^{2}a_{n}=0,} where $\alpha$ is the strength of the linear potential, e.g., the variation of the effective refractive index in the WA.
A new term $\alpha n \Psi$ to the left-hand side of Eq. (\ref{mainsimplified}) will appear due to the linear potential
\eq{mainsimplifiedLP}{\left[ i\de_{z} +\alpha n - \frac{D_{2}}{2}\de_{n}^{2} + \sum_{m\geq3}\frac{D_{m}}{m!}(-i\de_{n})^{m} +  |\Psi|^{2}\right]\Psi=0.}
By using the moment method developed to calculate the rate of SSFS for fibers \cite{agrawal2,santhanam} the wavenumber in SSWS can be calculated as follows:
\eq{SR}{\kappa(z) = \frac{-i}{2E}\int_{-\infty}^{+\infty} \left(\Psi^{*}\de_{n}\Psi - \Psi\de_{n}\Psi^{*}\right)dn,} where $E = \int_{-\infty}^{+\infty}|\Psi|^{2}dn$ is the total energy. From Eqs.(\ref{mainsimplifiedLP}) - (\ref{SR}) the {\em soliton self-wavenumber shift} (SSWS) rate turns out to satisfy the simple expressions for the central wavenumber
\eq{fSR}{\kappa(z) = \kappa_{0} + \alpha z,}
and the central position of the soliton
\eq{nSR}{n_{0}(z)=n_{0}(0)+c \mathrm{cos}(\kappa_{0})(\alpha z^{2} + 2\kappa_{0}z).}
Thus, the wavenumber shift rate is simply equal to $\alpha$, which is also proportional to the rate of acceleration or deceleration [for $\sgn(\alpha)=\pm1$, respectively] in the $n$ space.

Actually the above formula of $\kappa(z)$ is exact in the absence of Kerr nonlinearity. One immediate effect of the wavenumber shift in the {\em linear} regime is the onset of the well-known photonic Bloch oscillations (PBOs), see Refs. \cite{peschel,pertsch2,lenz,lederer}. We can interpret PBOs in the following way. In Fig. \ref{fig2}(a) a linear beam is launched  in a WA made of $N=35$  waveguides at normal incidence ($\kappa_{0}=0$), but due to the wavenumber shift described above, its wavenumber changes linearly towards positive detunings $\Delta\kappa$ (for $\alpha>0$), and therefore it accelerates towards positive values of $n_{0}$, as shown in Fig. \ref{fig2}(a). However, due to the folding of the band structure, when the wavenumber reaches the value $\kappa=\pi$, it folds back into the first Brillouin zone, and thus $\kappa$ becomes negative, see white solid lines in Fig. \ref{fig2}(b). This means that the beam must change direction, and starts decelerating towards negative values of $n_{0}$, and this process is repeated indefinitely in a sinusoidal fashion with a period $z_{0} = 2\pi/\alpha$, see Fig. \ref{fig2}(a). Therefore, the above formula for $n_{0}(z)$ is only valid at the beginning of propagation, before the occurrence of the folding effect.

The analogy between SSFS and SSWS appears in all its clarity in the {\em nonlinear} regime. Even though the soliton input wavenumber is far from the zero diffraction points (located at $\kappa=\pm\pi/2$, see Refs. \cite{lederer,truong2} for more details), eventually the SSWS will push the soliton wavenumber in proximity of one of these two points, depending on the sign of $\alpha$. At this moment, one can conjecture that the soliton will start emitting a strong DifRR that is able to compensate the  SSWS, in exactly the same way how solitons emit strong DisRR when approaching a zero-GVD point in photonic crystal fibers \cite{skryabinscience}. This is indeed the case, as shown in Figs. \ref{fig2}(c,d). In Fig. \ref{fig2}(c) we show the $z$ evolution of a soliton in the $n$ space with an initial position $n_{0}(0)=0$, in a potential with $\alpha=0.01$. We use a WA made of $N=300$ waveguides, but this is done only in order to make all the figures as clear as possible for the reader, while the concept works for an almost arbitrary number of waveguides (typically $\sim 10$ waveguides would suffice). The soliton starts accelerating (for $\alpha>0$) towards positive values of $n_{0}$ or decelerating (for $\alpha<0$) towards negative values of $n_{0}$ continuously, until at a specific moment (around $z\simeq 70$) it emits the DifRR. In Fig. \ref{fig2}(d) we show the same as Fig. \ref{fig2}(c), but in the wavenumber space. The soliton starts with zero central wavenumber (i.e. normal incidence, $\kappa_{0}=0$), shifts continuously and linearly its wavenumber (the white solid line shows the predicted shift, $\kappa(z) = \kappa_{0} + \alpha z$, for the central wavenumber) until a considerable amount of soliton energy reaches the zero diffraction point $\kappa=\pi/2$, which triggers the emission of a strong DifRR. This emission is so strong that the wavenumber shift of the soliton is compensated, a process that is an exact analogue of the SSFS compensation in fibers \cite{biancalana,skryabinscience}.

However, note that in Fig. \ref{fig2}(c,d), an anomalous recoil, which has been described in Ref. \cite{truong2}, occurs: the DifRR should be emitted by the soliton in the positive detunings $\Delta\kappa>0$, but the folding of the Brillouin zone makes the radiation appearing in the negative detunings -- and thus in this case also the soliton central wavenumber shifts towards the DifRR, and not away from it, as it happens in optical fibers.

\section{Conclusions and future work}
\label{conclusions}
In conclusion, we demonstrate that it is possible to mimic closely temporal fiber-optical dynamics, unveiling the
new effects of wavenumber-supercontinuum generation and the compensation of the 'soliton self-wavenumber
shift' by the emitted diffractive resonant radiation. Because of the periodicity, several new effects can take place, which have no counterpart in continuous systems, such as the anomalous soliton recoil and anomalous compensation of the soliton self-wavenumber shift. This work paves the way for designing unique optical devices that generate spectrally broad supercontinua with a controllable directionality and many other spatiotemporal effects in waveguide arrays. Moreover, this analysis is applicable to virtually any nonlinear discrete periodic system supporting solitons, therefore making our results very general and of relevance for a number of very diverse communities.

Future works include the investigation of the SCG both in frequency and wavenumber, and also the generation of light bullets in WAs.

FB (Max Planck Research Group) and TXT (Max Planck Partner Group) are supported by the German Max Planck Society for the Advancement of Science (MPG).

\end{document}